\def\preprint{0}		%  submitted version i.e single col 
\def\preprint{1}		%  preprint i.e. MNRAS style

\if\preprint1
	\documentclass{article}
	\usepackage{mn2e}
	\usepackage{times}
	\usepackage{graphicx}
	\usepackage{calc}

\else
	\documentclass{article}
	\usepackage[referee]{mn2e}
	\usepackage{times}
	\usepackage{graphicx}     % inserted by JGR - to get figs in ref vers
	\usepackage{calc}         % inserted by JGR
\fi

\newcommand{\comment}[1]{\relax}

\ifnfsstwo

\fi

\ifnfssone
  \newmathalphabet{\mathit}
    \addtoversion{normal}{\mathit}{cmr}{m}{it}
    \addtoversion{bold}{\mathit}{cmr}{bx}{it}

\fi

\ifoldfss

\fi

\loadboldmathitalic
\loadboldgreek

\title[Sight line towards Q2059$-$360]{Emission within a Damped
Lyman~$\balpha$
Absorption Trough: the Complex Sight Line Towards Q2059$-$360\thanks{Based on
observations obtained at the ESO La Silla Observatory}}

\author[B.~Leibundgut and G.~Robertson]{
Bruno Leibundgut$^1$ and
J.~Gordon Robertson$^{2,1}$\\
$^1$European Southern Observatory, Karl-Schwarzschild-Strasse 2, D-85748
Garching, Germany \\
$^2$School of Physics, University of Sydney, NSW 2006, Australia}

\date{Accepted .... Received .....}
\pagerange{\pageref{firstpage}--\pageref{lastpage}}
\pubyear{1998}

\begin{document}

\maketitle

\begin{abstract}

We present new spectroscopic observations of the quasar Q2059$-$360,
confirming the existence of an emission feature within the Damped Lyman
Alpha (DLA) absorption trough. By observing also at slit positions offset
from the quasar, we show that the emission is spatially extended by at
least a few arcseconds, and hence confirm that the feature seen must be due
to emission rather than unusual absorption characteristics.  We find that
the DLA trough is very close in redshift to the broad Lyman~$\alpha$
emission line of the QSO, with the result that the DLA absorption removes
much of the peak region of that line.  Despite the similarity of the
redshifts of the DLA and the QSO, the lack of high-ionization lines of the
DLA system and the unresolved widths of the corresponding metal lines
indicate that the DLA cloud is not an associated system.

The emission feature has a large velocity offset of +490 km~s$^{-1}$ with
respect to the DLA system, and is resolved in velocity, comprising two
components with a separation of $\sim$ 300 km s$^{-1}$.  We consider three
possibilities: (1) Both emission and absorption occur within an object
similar to the high redshift Lyman-break galaxies; (2) The emission feature
arises from an object distinct from both the DLA absorber and the QSO,
perhaps a young star-forming galaxy or a proto-galactic clump. It could be
associated with the DLA absorber and perhaps the QSO in a compact group or
cluster; (3) The redshifts are such that the emission feature could be due
to Narrow Line Region filaments of the QSO, if the DLA absorption
covers a sufficiently small angular size to allow the filaments to be seen
beyond the edge of the DLA cloud.

\end{abstract}


\section{Introduction}

The systems causing redshifted Damped Lyman Alpha (DLA) absorption features
in the spectra of quasars are the subject of many present studies, largely
because they may be the progenitors of today's normal galaxies, and because
they represent the dominant form in which neutral hydrogen occurs at redshifts
\raisebox{0.5ex}{$>$}\hspace*{-0.7em}\raisebox{-0.7ex}{$\sim$} 2.  The DLA
column densities are usually taken to be $N$ ({H\ {\sc i}})
\raisebox{0.5ex}{$>$}\hspace*{-0.7em}\raisebox{-0.7ex}{$\sim$} 10$^{20}$
cm$^{-2}$, typical of the column in a present-day disc galaxy.  Selection
by such passive absorption is more likely to favour `normal' galaxies, as
opposed to those which may be found using selection based on the presence
of an AGN.  The similarity of column densities does not itself establish
that DLAs are due to galactic discs, but some further support has come from
the disc-like kinematics of DLAs observed by Prochaska \& Wolfe (1996),
although Haehnelt et al. (1997) show that merging pre-galactic clumps can
also explain the observations. A connection between DLAs and galaxies is
also suggested by the near equality of the high-redshift total mass density
of DLAs and the measured total mass in galaxy discs in the local universe
(Storrie-Lombardi et al. 1996). On the other hand, it is becoming apparent
that there is a wider diversity of objects causing DLA systems at
intermediate redshifts than for the
intervening Mg~{\sc ii} absorption systems (Le Brun et al. 1997, Steidel et
al. 1997a). The possible selection effect caused by dust extinction in
discs of chemically evolved spiral galaxies must also be considered (e.g.
Steidel et al. 1997a). 

Absorption spectra examine a DLA absorber only along a single sight line,
giving no direct indication of spatial extent and hence total H~{\sc i}
mass. Additional insight into the nature of the DLA systems could be gained
by detection of their emission, which should be feasible if they are indeed
of galactic dimensions. A protogalaxy or young `normal' disc galaxy can be
expected to have a significant population of young stars, which will excite
the surrounding gas and result in a variety of emission lines, including
Lyman~$\alpha$.

Several searches for such Lyman~$\alpha$ emission have been made, but with
limited success.  Hunstead, Pettini \& Fletcher (1990) reported detection
of emission from the $z_{\rm abs}$ = 2.465 DLA towards Q0836+113. The DLA
feature totally blocks the light from the background QSO over a range of
some 15~\AA. Using intermediate dispersion spectroscopy, Hunstead et
al. were thus able to see the faint emission from the DLA `galaxy' against
a dark background rather than superimposed on the QSO continuum. The result
was confirmed with independent observations at a different telescope
(Pettini et al. 1997), although other attempts to confirm this detection by
use of narrow band filter imaging and further spectroscopy were
unsuccessful (Wolfe et al. 1992, Lowenthal et al. 1995).  A few other
detections have been reported: PKS 0528$-$250 at $z_{\rm abs}$ = 2.811 (M\o
ller \& Warren, 1993; Warren \& M\o ller, 1996), Q2059$-$360 at $z_{\rm
abs}$ = 3.0831 (Pettini et al. 1995), Q2233+131 at $z_{\rm abs}$ = 3.1501
(Djorgovski et al. 1996) and Q0151+048A at $z_{\rm abs}$ = 1.9342 (Fynbo et
al. 1997, M\o ller et al.  1998).  The lack of detection in other cases,
and the weak line fluxes of the positive detections, have led many authors
to suggest that the Lyman~$\alpha$ photons are destroyed by dust in the
clouds, with the probability of absorption enhanced by resonant scattering
of Lyman~$\alpha$ photons (e.g. Charlot \& Fall, 1991).

Apart from giving detections of the DLA `galaxies'
the Lyman~$\alpha$ emission line flux is important in that it
can be used to indicate the star formation rate (SFR). Unless dust is
destroying much of the Lyman~$\alpha$ flux, the values (or limits) found
indicate surprisingly low SFRs, around 1 M$_{\odot}$ per year (e.g.
Hunstead et al. 1990).

We have commenced an observational project aimed at further clarification
of the status of the DLA `galaxies'. Despite the known difficulties, we
have sought to detect Lyman~$\alpha$ emission from these objects. In this
first paper we report confirmation of the emission from the DLA towards
Q2059$-$360, and further elucidation of its properties.  We have used
intermediate dispersion spectroscopy (rather than narrow band imaging)
because we believe it is essential to have adequate spectral resolution to
separate a claimed DLA emission feature from the QSO's light.  It is also
significant that the DLA's emission peak need not coincide spatially with
the background quasar; offsets of order 1$\arcsec$ are likely, given the
expected size of DLA absorbers (if they are indeed comparable in size to
galaxies).  Narrow band imaging is sensitive to objects anywhere near the
QSO (although with reduced sensitivity very close to the QSO) while in this
work we took spectra offset from the quasar position as well as centred on
it.  With the long-slit spectrograph, this provides a rudimentary
two-dimensional mapping capability, while retaining the full spectral
resolution. In the case of Q2059$-$360, this strategy has demonstrated that
the DLA emission is spatially extended, showing conclusively that it is a
separate emission object or region, and not a peculiarity in the absorption
of the background (spatially unresolved) QSO.  We discuss a number of
intriguing aspects of this system. The interpretation is complicated by the
very near equality of the QSO emission and DLA absorption redshifts (closer
than stated by Pettini et al. 1995), which gives rise to additional
possibilities for the origin of the DLA emission feature and its relation
to the absorber.

In the following we present our observations and the data reduction (\S~2).
The results regarding the quasar spectrum, the absorber and the excess
emission within the DLA trough are summarized in \S~3 which is followed by
a discussion of the geometry and the possible relations between QSO,
absorber and the emission (\S~4). We conclude in \S~5.

\section{Observations and Reductions}

The quasar Q2059$-$360 (RA 21 02 44.6, Dec $-$35 53 07, J2000) was discovered
by Warren et al. (1991). We observed it on the nights of 29, 30 and 31
August 1995 with the NTT at the ESO La Silla Observatory. The blue arm of
the ESO Multi-Mode Instrument (EMMI) was used to cover the wavelength range
from 4650 to 5100~\AA\ with a 1200~g~mm$^{-1}$ grating providing a
reciprocal dispersion of 17.5~\AA~mm$^{-1}$. The slit width was 1.2$\arcsec$
for all integrations, resulting in a resolution of 1.5 \AA\ FWHM measured
from lines in the comparison lamp spectrum.  A thinned Tektronix
1024$\times$1024 CCD binned in the spatial direction to yield a scale of
0.74$\arcsec$ per pixel was used.  Measurements of the spatial profile of the
quasar in a continuum band near the Lyman~$\alpha$ emission showed that the
spatial binning was justified, with  only a few observations being slightly
undersampled (cf. Table~1).
We also binned the CCD in the spectral direction to a reciprocal dispersion
of 0.84 \AA\ per pixel.  Although this undersampled the line profiles
slightly, it assisted in detecting any faint unresolved flux by
concentrating as much light as possible into a single pixel. The CCD
readout noise was 4.3 electrons rms.

Integrations of one hour each were obtained with the slit oriented
east-west and positioned at various declination offsets from the quasar to
map the spectrum of any extended or displaced emission source.  Individual
observations were also offset along the slit to position the spectrum at
different places on the CCD, so guarding against faint CCD artefacts
masquerading as signal. The observation log is presented in Table~1.

\begin{table*}
\caption{Observing Log for Q2059$-$360}
\begin{tabular}{l l l l r c l}
\hline
obs number & \multicolumn{2}{c}{date} & airmass & offset$^a$ & spatial & comments \\
 & \multicolumn{2}{c}{(UT Start)} & & \multicolumn{1}{c}{($^{\prime\prime}$)} & FWHM ($^{\prime\prime}$) & \\
\hline
49  & 1995 August & 29.013 & ~~1.19 & $-$0.4~~ & 1.9 & thin cirrus \\
50  &             & 29.060 & ~~1.06 &  0.0~~   & 1.3 & thin cirrus \\
56  &             & 29.125 & ~~1.01 &  1.7~~   & 1.7 & thin cirrus \\
71  &             & 29.227 & ~~1.22 &  1.0~~   & 1.6 & thin cirrus \\
120 &             & 29.981 & ~~1.33 & $-$0.5~~ & 1.4 & - \\
121 &             & 30.024 & ~~1.14 & $-$1.2~~ & 1.5 & affected by cosmic ray \\
124 &             & 30.072 & ~~1.04 & $-$1.9~~ & 1.3 & - \\
125 &             & 30.122 & ~~1.01 & $-$0.4~~ & 1.3 & affected by cosmic ray\\
184 &             & 30.993 & ~~1.25 &  0.0~~   & 1.5 & affected by cosmic ray\\
185 &             & 31.036 & ~~1.10 & $-$1.9~~ & 1.7 & - \\
188 &             & 31.085 & ~~1.02 & $-$2.3~~ & 1.7 & - \\
189 &             & 31.127 & ~~1.02 & $-$2.1~~ & 1.7 & - \\
\hline
\end{tabular}

\medskip

{\bf Note:} $^a$
`Offset' refers to the slit positioning with respect to the quasar;
positive offsets place the slit north of the quasar.
The determination of the offsets in declination was
complicated by errors in the recorded telescope positions; the values given
are accurate to $\sim 0.\arcsec 5$ --- $0.\arcsec 8$, well within the
slit width.

\end{table*}

The reductions included the removal of CCD instrumental effects. In this
process a low-level fixed pattern noise, probably from the CCD controllers,
was seen in the average of multiple bias or dark frames.  The amplitude
($\sim$2~$e^-$) was, however, negligible compared to the Poisson
fluctuations in the signal obtained from the sky background, which had an
average level of $\sim$100 $e^-$.  The target frames were then flatfielded
with high-level dome flat fields. Cosmic ray removal was performed manually
around the spectrum, because we are looking for emission with unknown width
and strength. However, at separations of more than about 10$\arcsec$ from
the quasar spectrum automatic removal of cosmic rays was used. We then
removed any background light by fitting polynomials to the smooth part of
the spatial intensity distribution (i.e. excluding the QSO spectrum) and
subtracting this background.  In this way the sky background was assessed
over a large number of spatial rows, with the result that sky subtraction
added negligible noise to the desired spectrum.  The smooth component of
the dark current was also removed by this process.  The wavelength solution
was found for each spectrum using ThAr lamp observations obtained directly
before or after each integration. The accuracy achieved is 0.22 \AA\ rms,
which is acceptable considering the pixel scale and the undersampling of
the spectrum.  The alignment and possible image distortions were determined
using standard star spectra and corrections applied to the two-dimensional
data at the same time as the wavelength transformation.  All spectra were
further corrected to vacuum wavelengths and transformed to the heliocentric
frame.

Flux calibration was performed using the standard stars LTT~7987 and LTT~9491
(Hamuy et
al. 1994) observed under the same conditions as the QSO, including a
comparable airmass.  The absolute flux zero-point varied by less than 15\%
for individual observations in the three nights.
We take this to be the uncertainty in the absolute flux
measurement. The relative flux calibrations as a function of wavelength
agree to better than 4\% between observations.

Note that the absolute flux calibration is not needed for the detection of
any emission associated with the damped system. As will be shown below
(section 3) the uncertainty in the emission flux is dominated by the low
signal level of the detection.

The `spatial FWHM' measurements reported in Table~1 are from
Gaussian fits to the spatial profile of the longslit spectra.  The
instrument focus of the EMMI blue arm makes a significant contribution to
the width, since the seeing as measured from direct images was appreciably
better (0.8$\arcsec$ FWHM in R). We set the spectrograph to the optimum focus
for the {\it spatial} profile, after tests showed that this did not
coincide with the best spectral focus.

A severe problem is the high cosmic ray rate of the chip. The automatic
procedure which we employed away from the spectrum typically removed about
1400 events per 1 hour exposure. In a few cases we had cosmic ray events
close to or in the region of the damped line. Those events were very
carefully removed manually and the result checked to be sure not to
introduce any
artefacts into the data. In three cases, indicated in Table~1, we could not
reliably remove the cosmic ray near or on the position of the emission
reported here. The affected regions of those three spectra were flagged
(e.g. in Figure \ref{fig.multispectra}) and those runs were excluded
from the spectra of the emission feature itself (Figure \ref{fig.blobspectra}).

\section{Results}

\begin{figure}
\centerline{
\includegraphics[bbllx=12mm,bblly=18mm,%
bburx=198mm,bbury=253mm,%
width=\the\hsize]{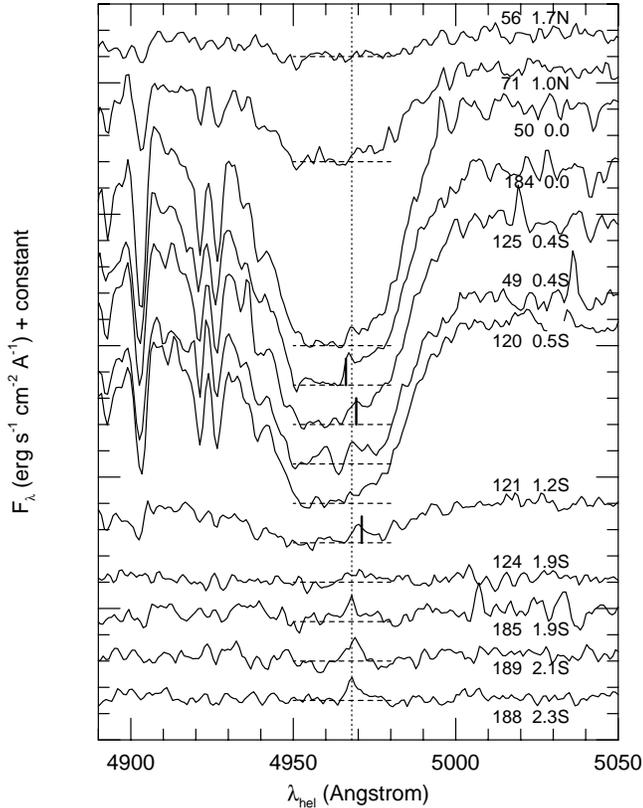}}
\caption[]{\label{fig.multispectra}
Individual observations near the sight line towards Q2059$-$360 identified
by their observation number and the offset position (in arcseconds). The
spectra are ordered according to the position relative to the QSO (from
North through South). One tick on the ordinate corresponds to
$2\cdot10^{-17}$~erg~s$^{-1}$~cm$^{-2}$~\AA$^{-1}$.  Wavelengths affected
by cosmic rays (or residuals of their removal) in the range of the damped
Lyman~$\alpha$ absorption are indicated by the heavy vertical strokes. The
horizontal dashed lines show the zero level for each spectrum, while the
vertical dashed line indicates the central wavelength of the emission
feature.  The spectra have been smoothed (see text).}

\end{figure}

The 12 spectra are displayed in
Figure \ref{fig.multispectra}, in order of declination offset. Only the
part of the spectrum near the DLA trough is shown, and the spectra have
been smoothed with a Gaussian of $\sigma =$ 0.7\AA\ to optimally show the
weak emission feature. The drop in intensity of the QSO spectrum at larger
offsets is clearly visible.  The scatter around the zero level in the four
spectra between 1.9 and 2.3 arcseconds south, where the QSO continuum is
negligible, indicates the noise level achieved by the individual
integrations.  The spectra centered on Q2059$-$360 clearly show the damped
system, and a few Lyman forest lines on the blue side of the trough.

\begin{figure*}
\centerline{
\includegraphics[%
bbllx=16mm,bblly=9mm,%
bburx=251mm,bbury=183mm,%
width=\the\hsize,height=112mm]{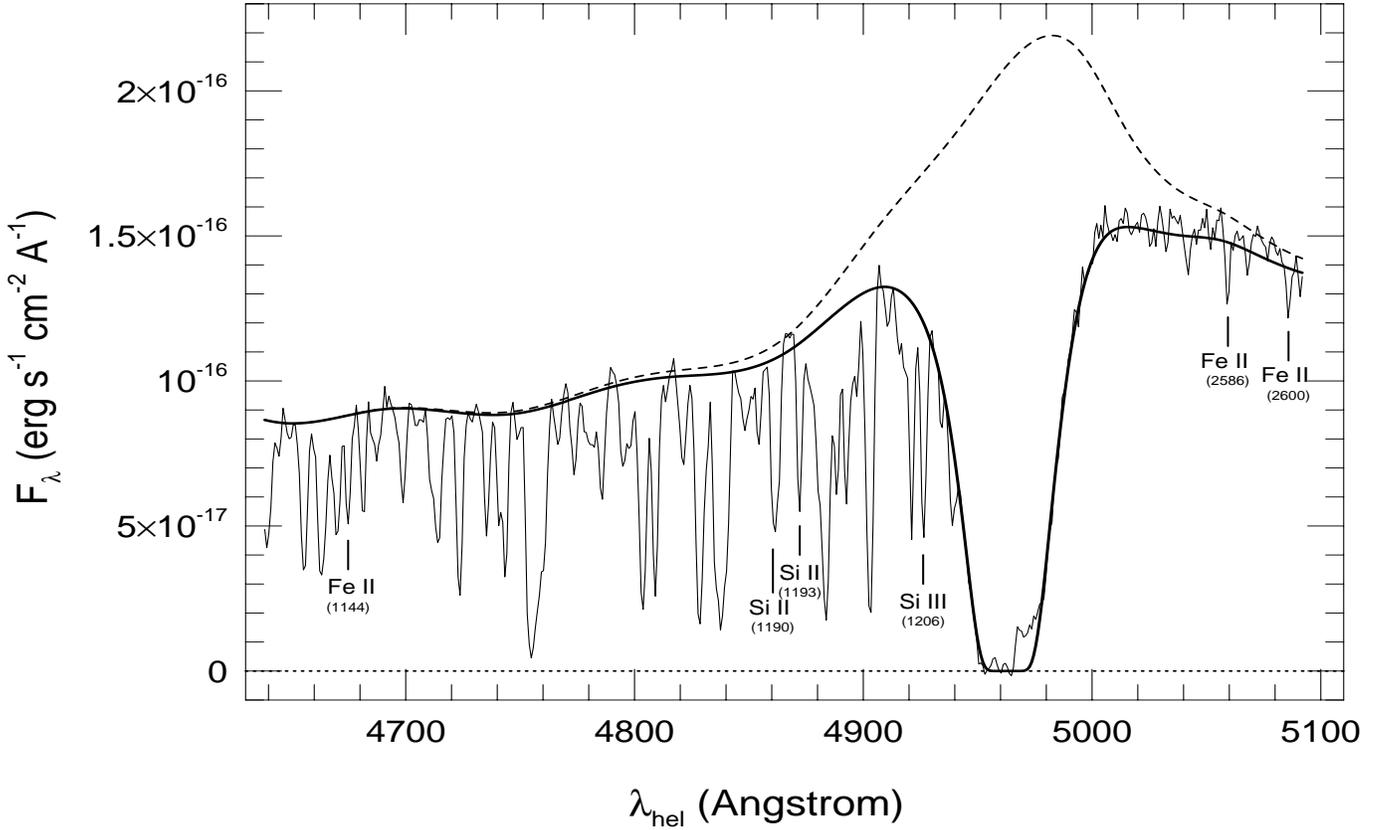}}
\caption[]{\label{fig.qsospectrum} 
Combined spectrum towards Q2059$-$360. The derived fit of the QSO
continuum, including the broad Lyman~$\alpha$ emission line, is shown by
the dashed line. The heavy continuous line shows the same QSO continuum
after absorption by the damped system as modeled (see text for
details). Also indicated are the wavelengths of metal lines for an assumed
redshift of 3.08303, and the two lines in a lower redshift system.}

\end{figure*}

For the investigation of the quasar spectrum we combined the five
observations with offsets smaller than 1$\arcsec$, totalling 5 hours. The result
is shown in Figure \ref{fig.qsospectrum}, which covers the full observed
wavelength range, and is unsmoothed.  As well as the DLA trough, the
spectral region observed covers several metal lines in the $z_{\rm abs}
\approx$ 3.083 (DLA) system.  Si~{\sc ii} 1190 appears blended with a Lyman
forest line, but Fe~{\sc ii} 1144, Si~{\sc ii} 1193 and Si~{\sc iii} 1206
give good detections and show no evidence of blending. Line fitting was
performed using the Xvoigt package (Mar \& Bailey 1995).  The redshifts
derived from these lines are 3.08294, 3.08299 and 3.08317, respectively,
giving a mean of 3.08303 $\pm$ 0.00007, in good agreement with the value
derived by Pettini et al. (1995).  The metal lines are essentially
unresolved in our intermediate dispersion spectra, giving an upper limit to
the Doppler velocity dispersion parameter $b$ of 60 $-$ 80 km s$^{-1}$ and
a best estimate of 30 km s$^{-1}$.  Accurate column densities cannot be
obtained because our spectral resolution is not adequate to rule out
saturation of the lines.  Nevertheless, the observations indicate iron and
silicon abundances of $\sim$0.015 solar and $\sim$0.04 solar respectively,
if we use the best estimate $b$ value and the N(H~{\sc i}) value given
below.  If confirmed by higher resolution observations, the results show
the low enrichment typical of DLAs and high redshift intervening QSO
absorption systems in general (Lu et al. 1996, Pettini et al. 1997). No
other metal lines in this absorption system were detected in our spectrum,
largely due to the restricted wavelength coverage and the effects of the
Lyman~$\alpha$ forest. The N~{\sc v} 1238, 1242 lines are not detected, to
a limit of 0.07 \AA\ (3 $\sigma$) equivalent width in the rest frame.
There is also a probable absorption system at a redshift of 0.9560 $\pm$
0.0001, which produces Fe~{\sc ii} 2586, 2600 lines longward of the
Lyman~$\alpha$ emission peak.

Figure \ref{fig.qsospectrum} also shows the prominent DLA absorption. In
order to estimate the neutral hydrogen column, and to optimally separate
the DLA-emission feature from the QSO emission, we fitted a Voigt profile
to the DLA, again using Xvoigt. The process of continuum fitting was
complicated by the proximity of the DLA absorption to the broad
Lyman~$\alpha$ emission line of the QSO. We performed a quasi-simultaneous
fit to the absorption and QSO emission profiles, via the following steps:
(a) A preliminary continuum fit, bridging smoothly across the DLA; (b) A
preliminary fit to the DLA (leaving large residuals in both wings of the
line, due to the inappropriate continuum); (c) Formation of a spectrum
equal to the original spectrum divided by the model DLA fit, the latter
truncated below 0.025 transmission. This gives an estimate of the quasar
spectrum without the effects of the DLA; (d) A smooth fit to the latter
spectrum, bridging across the noisy range where the DLA is black and/or
affected by the emission feature.  This provides a new estimate of the
continuum + emission line spectrum of the quasar; (e) A new fit to the DLA,
etc. The procedure was iterated 3 times, with only minor corrections after
the second pass.  The procedure works because it is able to make use of the
{\it sides} of the damped profile, where transmission is measurable and is
influenced by both the emission and absorption profiles. The truly black
part of the DLA is sufficiently narrow that it can be successfully bridged
in step (d), provided that we accept that the emission profile should vary
smoothly.

Figure \ref{fig.qsospectrum} includes the resulting fit and shows that the
DLA has removed the major Lyman~$\alpha$ emission peak of the QSO.  We
measure an emission redshift for the model of 3.097 $\pm$ 0.005 which is
closer to the original $z_{\rm em} =$ 3.09: value given by Warren et
al. (1991) than to the $z_{\rm em} =$ 3.13 value given by Pettini et
al. (1995).  Inspection of the emission spectrum given by Warren et
al. (1991) and a low dispersion spectrum obtained by us (Leibundgut and
Robertson 1998) shows that the redshift based on C~{\sc iv} 1548, 1550
alone is 3.085 (using $\lambda_{\rm eff, vac} =$ 1549.06\AA, Gaskell
1982). Moreover, the equivalent width of the Lyman~$\alpha$/N~{\sc v}
emission line is only half that of C~{\sc iv}, whereas it is normally about
twice the strength of C~{\sc iv}, again consistent with the DLA having
removed the major part of the Lyman~$\alpha$ emission line.

For the DLA itself we measure a column density of $\log(N) =$ 20.85$\pm$0.03.
Both wings of the damped profile are fitted well.  The redshift of the DLA
is $z_{\rm abs} =$ 3.0825. The fact that we find a slightly higher column
density compared with the Pettini et al. (1995) value of $\log(N) =$ 20.70
can be attributed to the different treatment of the broad QSO emission line
and a small but significant redshift difference between the DLA absorber
and the metal lines in the same system. We obtain a better fit of the model
DLA/continuum to the data than that shown by Pettini et al., who
constrained their fit to the redshift derived from the metal lines.  The
metal lines give a redshift about 40 km s$^{-1}$ larger than that of the
best fit Voigt profile for the DLA itself. This discrepancy is somewhat
larger than the uncertainty we expect even with the complication of fitting
the QSO's broad Lyman~$\alpha$ emission line superimposed on the DLA.

\begin{figure}
\centerline{
\includegraphics[%
bbllx=57mm,bblly=180mm,%
bburx=149mm,bbury=257mm,%
width=82mm]{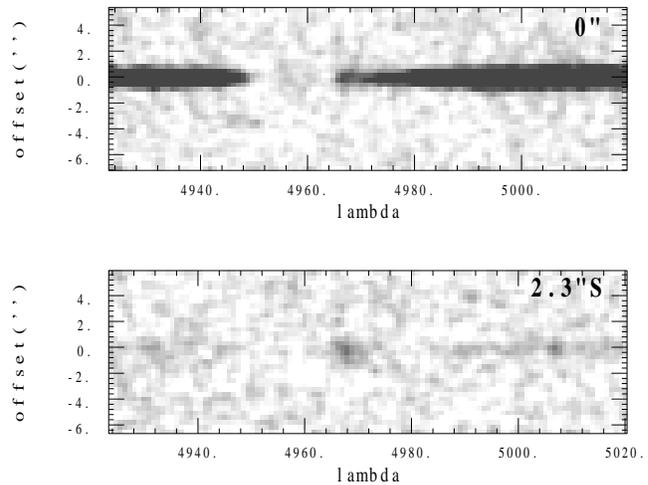}}
\caption[]{\label{fig.2dspectra} Two-dimensional spectra of the region
around the DLA. For the upper panel
all spectra near the QSO have been combined, while for the lower panel
all observations near offset
2$^{\prime\prime}$S are included.
The wavelengths shown are vacuum heliocentric; East is
up. Both images are displayed with the same greyscales.
Two
dimensional Gaussian smoothing has been applied, with sigma of 0.70\AA\ in
the wavelength and 0.\arcsec51 in the spatial directions.}

\end{figure}

Excess emission is clearly detected at the bottom of the damped line.  As
Figure \ref{fig.multispectra} shows, this emission is detected in several
individual exposures including positions offset from the QSO sight line
(spectra 124, 185, 188, and 189).  All spectra obtained on or South of the
quasar show excess emission at $\lambda_{\rm hel} \sim$ 4968\AA, although
in some cases rendered unreliable by nearby cosmic ray hits.  To help
illustrate the reality of the detections we show in Figure
\ref{fig.2dspectra} the combined longslit spectra of all observations on
the QSO and the 4 observations at an offset of 2$^{\prime\prime}$ South,
smoothed to match the instrumental resolution in the spectral direction and
the profile width in the spatial direction.  In all images except the
northern positions we clearly detect the emission at the bottom of the
Lyman~$\alpha$ trough, demonstrating the reality of the emission.

\begin{figure}
\centerline{
\includegraphics[%
bbllx=11mm,bblly=14mm,%
bburx=255mm,bbury=185mm,%
width=\the\hsize]{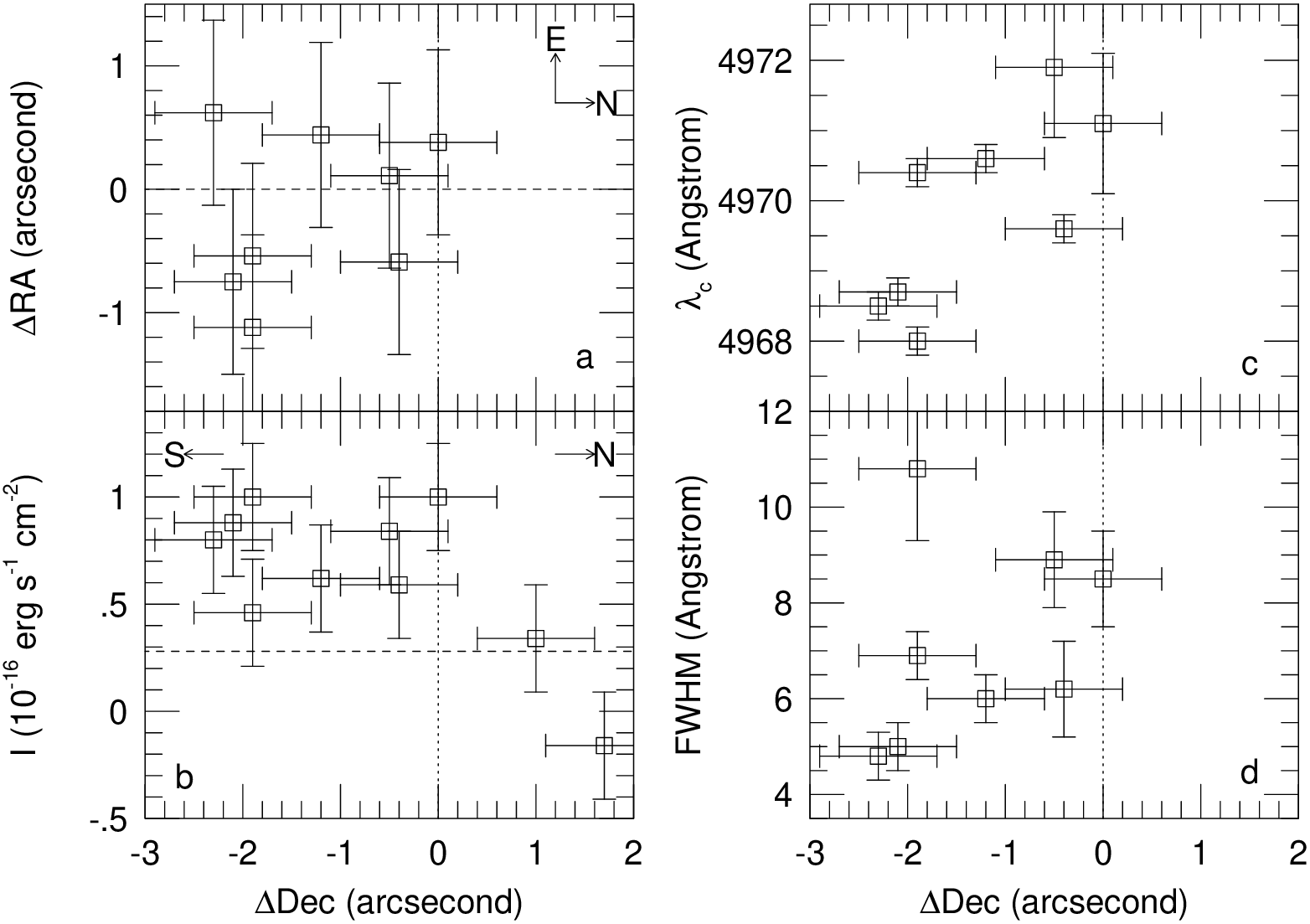}}
\caption[]{\label{fig.4plots}
Variation of properties of the emission feature as a
function of the NS offset of the slit in different runs. Observations
in which the emission feature was undetectable, or was affected by a cosmic
ray near the emission feature, have been omitted. All measurements were
made on the smoothed data.
(a) Centroid position of
the emission feature along the slit; (b) Flux of the emission feature (the
horizontal dashed line shows the 1$\sigma$ flux above zero);  (c) Centre
wavelength, derived from Gaussian fit; (d) Width of the emission feature.}
\end{figure}

In order to examine any spatial separation of the weak DLA emission from
the QSO in the RA direction (i.e. along the slit), we determined the
exact location of the quasar spectrum in the RA direction by collapsing the
data in the spectral direction in the range above 5000\AA, where the quasar
continuum dominates. This is reliable for all observations where we do
detect quasar light, but could be misleading for the largest offset
positions.  The separate emission feature was also located by a Gaussian
centering algorithm and the position relative to the QSO is recorded in
Table~2.  The resulting East-West positional displacement of the DLA
emission feature with respect to the QSO is shown in Figure
\ref{fig.4plots}a.  To within the uncertainties the emission remains
centred on the QSO in this direction for all the observations at various
N-S positional offsets.

\begin{table*}
\caption{Parameters of the emitter}
\begin{tabular}{l c c c c}
\hline
obs number & wavelength & displacement$^a$ (East-West) & Ly~$\alpha$ flux
& FWHM \\
& \multicolumn{1}{c}{(\AA)} & (\arcsec) & ($10^{-16}$ erg s$^{-1}$ cm$^{-2}$)
& \AA \\
\hline
49  & 4969.6  &   0.59    & 0.59 & 6.2 \\
50  & 4971.1  &   0.38    & 1.0  & 8.5 \\
56  & $\ldots^c$ & $\ldots$ & $-$0.16 & $\ldots$ \\
71  & $\ldots^c$ & $\ldots$ & 0.34  & $\ldots$ \\
120 & 4971.9  &   0.11    & 0.84 & 8.9 \\
121 & 4970.6$^a$ &  0.44    & 0.62 & 6.0 \\
124 & 4970.4  &  $-$0.54    & 1.0  & 10.8$^d$ \\
125 & $\ldots^b$ &$\ldots$ & $\ldots$ & $\ldots$  \\
184 & $\ldots^b$ &$\ldots$ & $\ldots$ & $\ldots$  \\
185 & 4968.0  &  $-$1.12    & 0.46  & 6.9 \\
188 & 4968.5  &   0.62    & 0.80  & 4.8 \\
189 & 4968.7  &  $-$0.75    & 0.88  & 5.0 \\
\hline
\multicolumn{4}{l}{Notes: $^a$`Displacement' refers to the spatial
separation of the emission}\\
\multicolumn{4}{l}{~~~~~~~~~~~~~~~from the quasar spectrum, with East offsets
positive.}\\
\multicolumn{4}{l}{~~~~~~~~~~~~ $^b$affected by nearby cosmic ray} \\
\multicolumn{4}{l}{~~~~~~~~~~~~ $^c$line not detected} \\
\multicolumn{4}{l}{~~~~~~~~~~~~ $^d$diffuse appearance in 2D spectrum} \\

\end{tabular}

\end{table*}

\begin{figure}
\centerline{
\includegraphics[%
bbllx=31mm,bblly=18mm,%
bburx=255mm,bbury=191mm,%
width=\the\hsize]{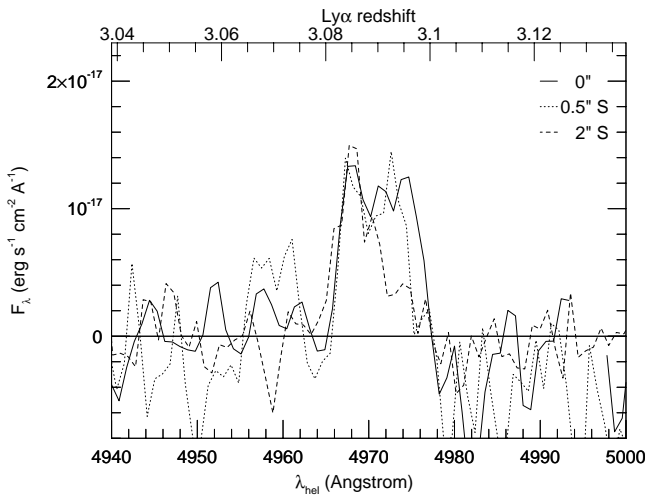}}
\caption[]{\label{fig.blobspectra} Spectra of the emission feature,
isolated from the QSO spectrum by subtraction of a scaled QSO
spectrum. Spectra have been averaged in three groups as a function of NS
offset of the slit from the quasar, but omitting runs affected by cosmic
rays near the emission feature. As a result, a single run (50) is shown for
the on-QSO position, two runs (120, 49) were averaged for offsets near
0.5$\arcsec$ S of the quasar, and four runs (124, 185, 189, 188) for offsets
near 2$\arcsec$ S of the QSO.  Spectra are unsmoothed, giving a resolution of
1.5\AA\ FWHM, except for the single on-QSO run, which was lightly smoothed
by a Gaussian of $\sigma =$ 0.7\AA.}
\end{figure}

The flux of that part of the DLA emission feature which passed through the
slit in each observation was measured by integrating over the wavelength
range of the emission after subtraction of the QSO's light. The background
level was taken from the wavelengths blueward of the emission feature,
i.e. in the black part of the DLA trough, with no contribution from
wavelengths redward of the emission feature, where residuals of the
subtraction process increase the uncertainties.  The resulting fluxes are
given in Table~2.  The error of these fluxes is about 3$\cdot10^{-17}$ erg
s$^{-1}$ cm$^{-2}$ for an integration interval of 12\AA, as measured from
the scatter in the background region. The fluxes are plotted in
Figure~\ref{fig.4plots}b. With the available signal-to-noise ratio there is
no clear trend of the total flux with offset position south of the QSO.

The centroid wavelength of the DLA emission feature was determined for each
observation (other than those affected by cosmic rays in the critical
places) and is plotted in Figure~\ref{fig.4plots}c. There is an indication
of a shift of the central wavelength with offset from the quasar. We
discuss below whether this can be interpreted as a rotation curve.

Finally, in Figure~\ref{fig.4plots}d we show evidence for narrowing of the
lines with increasing offset distance from the quasar. There is, however, a
strongly discrepant point in this relation from one spectrum which appears
to show a diffuse and possibly multi-component structure. With the S/N
available to us we cannot be certain of the reality of the narrowing trend.
As discussed below, the emission feature probably consists of two
components in velocity, and some narrowing in total velocity width is to be
expected as the higher velocity component fades at southern offsets, while
the flux of the lower velocity component is maintained (as shown in Figure
\ref{fig.blobspectra}).

Using the model fit of the damped system shown in
Figure~\ref{fig.qsospectrum} we subtracted the quasar's contribution from
all the spectra we obtained.  (The model spectra were scaled by the flux
observed between 5000 and 5080~\AA, where the QSO continuum dominates.)
This gave the best estimate of the spectrum of the emission feature as
isolated from the QSO/DLA spectrum. Figure \ref{fig.blobspectra} shows the
results, after averaging the observations in three groups with different
amounts of N-S offset from the quasar.

\section{Discussion}

\subsection{The Line of Sight Towards Q2059$-$360}
We now consider the nature of the DLA absorber and the Lyman~$\alpha$
emitter, and the possible relationships among these two objects and the
quasar. The large velocity difference between the absorber and the emitter,
and the relatively small velocity difference between the absorber and the
quasar emission may indicate a complex link between these three observed
components.  We will review the evidence and examine possible
interpretations of the data.

\subsection{Identification of the Rest Wavelength of the Emission Feature}
We first examine whether the emission feature should be interpreted as
Lyman~$\alpha$ emission. While it is not possible to be certain with the
present data, which show only the one emission line, we nevertheless
believe that a strong case can be made that the line is due to
Lyman~$\alpha$.

The only plausible alternative identification would be [O~{\sc ii}] 3726.1,
3728.8 at a redshift of 0.3333. The velocity splitting of these lines is
217 km s$^{-1}$. This is less than the observed separation of the two
emission components (approximately 300 km s$^{-1}$), but the limited
signal/noise of the data could still allow this identification. We base our
rejection of this hypothesis on the low probability of such a close
coincidence in {\it both} wavelength and position on the sky. If the object
is a foreground galaxy at $ z=$ 0.3333, then it is totally unrelated to
both the DLA and the QSO. The near coincidence in wavelength with the DLA
absorption is not in itself extremely unlikely, but combined with the {\it
very} close spatial alignment, the identification with the [O~{\sc ii}]
line is unlikely. It could be argued that we have effectively selected such
a system, by searching for emission near QSOs. But the search has been made
for only a small number of quasars, and the spatial alignment is much
closer than necessary for the emission to be detected (when we include
narrow band filter searches by other investigators).  In contrast, if the
line is Lyman~$\alpha$, then the spatial alignment follows from some form
of physical association between the emitter and the DLA and/or the QSO.  We
might also expect to detect continuum emission from the galaxy if it is at
$z =$ 0.3333. Should the emission indeed be the [O~{\sc ii}] doublet, then
the line ratio changes dramatically with spatial offset.  In fact, the
ratio at the largest offset position (2$^{\prime\prime}$ from the DLA) is
close to the optically thick value of 0.3 (Osterbrock 1989), implying the
highest density gas there.  The line ratio of unity (coincident with the
DLA) indicates a low density region there. This would represent a very
strong density gradient in the [O~{\sc ii}] over a range of about 12 kpc ,
although it would not be unreasonable for a normal galaxy.  (We use
$H_0=$ 50 km s$^{-1}$Mpc$^{-1}$). 

Further observations of the emitter to search for the C~{\sc iv} 1548, 1550
line (at $\lambda$ 6328) and for the continuum will be attempted. For the
present, we will treat the emission feature as representing Lyman~$\alpha$
emission.

\subsection{The absorber}
\begin{figure}
\centerline{
\includegraphics[%
bbllx=11mm,bblly=65mm,%
bburx=197mm,bbury=230mm,%
width=\the\hsize]{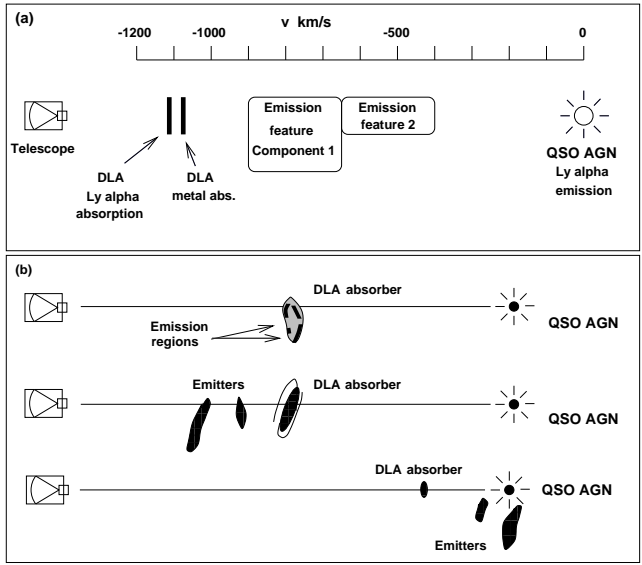}}
\caption[]{\label{fig.velpos2}
(a) Placement of the emission and absorption features on a radial velocity
scale. (b) Three possible spatial configurations for the Q2059$-$360 system
(not to scale).
}
\end{figure}

The DLA system is similar in H~{\sc i} column density to typical DLAs in
quasar spectra, but its velocity separation from the quasar emission
redshift is unusually small.  It is now well known that the broad QSO
emission lines Lyman~$\alpha$ and C~{\sc iv} are subject to mean blueshifts
and substantial scatter of their velocity with respect to the systemic
velocity of the QSO host galaxy (Gaskell 1982; Tytler \& Fan 1992).  With
only these two QSO emission lines observed (the C~{\sc iv} line from Warren
et al. 1991 or from Leibundgut and Robertson 1998), it is impossible to
calculate accurate velocities of the DLA absorption and emission features
with respect to the true QSO rest frame.  We have used the effective rest
wavelengths of Lyman~$\alpha$ and C~{\sc iv} from Tytler \& Fan (1992),
which at least remove the mean bias of redshifts calculated from these
lines.  The various emissions and absorptions are shown on a radial
velocity scale in Figure \ref{fig.velpos2}a, where the zero point for
velocity has been taken as the upper part of the peak of the QSO's
corrected Lyman~$\alpha$ emission line as reconstructed in Figure
\ref{fig.qsospectrum}.  The DLA absorption lies 1100 km~s$^{-1}$ blueward
of this, with the two emission feature components in between. However, if
the QSO redshift is taken instead from the peak region of the corrected
C~IV emission line, it falls only 500 km~s$^{-1}$ redward of the DLA,
i.e. within the second DLA emission feature.

While we cannot find a definite lower limit to the separation of the DLA
cloud from the QSO, it is clear that the DLA absorption system is not one
of the `associated' systems in close proximity to the parent QSO (Hamann
1997; Hamann et al. 1997). Such systems typically display high ionization
transitions and have relatively wide lines ($>$300 km s$^{-1}$). In
contrast, the damped system towards Q2059$-$360 does not display the N{\sc
v} doublet (see \S3), and the observed line widths are not more than about
100 km s$^{-1}$.  Associated systems also show metal abundances of solar or
higher (Petitjean \& Bergeron 1994, Hamann 1997, Franceschini \& Gratton
1997).  The abundances we calculated above ($\sim$ 0.015 solar and $\sim$
0.04 solar for iron and silicon respectively) are only approximate lower
limits, because of possible line saturation. But if the lines are indeed
sufficiently saturated for the abundances to approach solar values, then
the intrinsic velocity spread (Doppler $b$ parameter) can be only some 10
to 15 km s$^{-1}$, further increasing the contrast with the line widths
typical of associated systems.  Moreover, other DLA systems with $z_{abs}
\approx z_{em}$ have been reported to be spatially separated from the
quasar (M\o ller et al.~1998, Ge et al.~1997).

It is clear that the DLA absorber is not significantly affected by
radiation from the quasar; the absorber has properties similar to those
systems known to be independent objects well separated from their
background quasar. The conventional interpretation of such objects is as
galaxy discs (Wolfe et al. 1995) or proto-galactic clumps (Haehnelt et
al. 1997).

Three possible spatial configurations of the QSO, absorber and emitter are
shown in Figure \ref{fig.velpos2}b.  The velocity difference between the
QSO and the absorber would be consistent with them being members of a group
or a cluster.

Another interesting feature of the absorber is the small but probably
significant 40 km~s$^{-1}$ redshift of the metal absorption lines relative
to the hydrogen in the DLA. Velocity differences of this order between
Lyman~$\alpha$ forest lines and the corresponding C~{\sc iv} lines have
been noted previously (Womble et al. 1996), but it is more surprising to
find a difference between the hydrogen and low ionization species which can
co-exist with H~{\sc i}. The Lyman~$\beta$ absorption line in the DLA system
has been observed (Pettini et al. 1995) and shows a similar velocity
offset with respect to the metal lines (Hunstead, private communication).

\subsection{The Lyman~$\balpha$ emission feature}
\subsubsection{Velocity relationships}
We should like to know what physical system gives rise to the
Lyman~$\alpha$ emission feature in the DLA trough seen towards Q2059$-$360,
and whether the emitter is associated with the absorber, the quasar, or
neither.  Unlike the Ly~$\alpha$ emission features in PKS 0528$-$250
(Warren \& M\o ller 1996) and Q0836+113 (Hunstead et al. 1990), which are
centered in wavelength on the respective damped systems, the emission in
the absorption trough towards Q2059$-$360 is redshifted by about 490
km~s$^{-1}$.  Velocities of up to 300~km~s$^{-1}$ have been observed in
other detections of emission/absorption pairs (Warren \& M\o ller 1996,
Djorgovski et al. 1996, Lu et al. 1997).  

The large velocity difference between DLA absorption and emission rules out
the interpretation of both arising from a dwarf galaxy, as proposed for
Q0836+113 (Hunstead et al. 1990).  Even a large disc galaxy, with
absorption and emission occurring in different places in the disc
(Prochaska \& Wolfe 1997, Lu et al. 1997), is not likely to give an offset
of some 500 km~s$^{-1}$.  Instead, the results are more suggestive of
several pre-galactic clumps (Haehnelt et al. 1997; Warren \& M\o ller 1996;
Pascarelle et al. 1996).

However, recent work has shown that the velocity offsets between
Ly~$\alpha$ emission and stellar absorption lines observed in high-redshift
galaxies can amount to several hundred km~s$^{-1}$ (Franx et al. 1997,
Steidel et al. 1997b) or even exceed 1000 km~s$^{-1}$ (Pettini et
al. 1998). However, the highest values are thought to be partly due to
strong outflow in the interstellar medium, and it is not clear that a DLA
cloud could arise in such an outflow.   Velocity offsets of about
300~km~s$^{-1}$ between the Lyman~$\alpha$ emission and the galactic
absorption lines have also been observed in nearby galaxies and were
interpreted as due to infalling gas (Lequeux et al. 1995).

It is significant that our observations have resolved the emission feature
in both wavelength and spatial extent. Within the limited signal/noise of
the present data, it appears likely there are two components, as revealed
by the results shown in Figure \ref{fig.blobspectra}: component 1, with
emission centred at 4968.1 \AA\ ($z =$ 3.0867) and showing equal line flux
at all slit positions observed, both on and off the quasar; and component
2, centred at 4972.9 \AA\ ($z =$ 3.0907), showing equal line flux to
component 1 on the quasar and up to 0.5$\arcsec$ south, but tapering to a
much lower flux at 2$\arcsec$ south.  Our observation of emission component
1 corresponds to a physical extent of at least 14 kpc ($H_0=$~50 km
s$^{-1}$Mpc$^{-1}$, $\Omega=$~1) or 25 kpc ($H_0=$~50 km
s$^{-1}$Mpc$^{-1}$, $\Omega=$~0.1), i.e. of galactic dimensions. It is
remarkable, however, that one of the two emission line peaks shows no
perceivable wavelength shift over this spatial extent
(Fig.~\ref{fig.blobspectra}) .

As Figure \ref{fig.4plots}c shows, the centroid wavelength of the emission
feature shifts by about 3 \AA\ as it is traced from the QSO to the point
2$^{\prime\prime}$ away to the south.   If this Figure is interpreted as a
rotation curve, a lower limit on the mass of the system of 1.7 $\times
10^{10}$ M$_{\odot}$ ($\Omega =$~1) or 2.9 $\times 10^{10}$ M$_{\odot}$
($\Omega =$~0.1) can be derived. On the other hand, Figure
\ref{fig.blobspectra} makes clear that the shift of average wavelength
with declination is the result of a changing intensity ratio between the
two emission components, each of which has a constant wavelength. Thus the
mass limit may not be physically meaningful.

\subsubsection{Physical interpretation}
The observed flux from the emission feature is comparable to some of the
Lyman~$\alpha$ emission lines observed in galaxies at $z=$ 3 (Steidel et
al. 1996, Lowenthal et al. 1997) and the overall spatial extent is similar
to the typical galaxy sizes at these redshifts (Giavalisco et al. 1996).
But we could be observing two distinct, point-like, emissions which are
separated by 1 - 2 $\arcsec$, or an extended source.  One interpretation of
our data is that we have observed a young star-forming galaxy or
proto-galactic clump(s) in close association with the galaxy causing the
DLA absorption.

In order to be seen at the wavelengths where the DLA is black, the emitter
must naturally be in front (or to the side) of the DLA absorbing cloud.
(The somewhat {\it higher} redshift of the emitter compared to the absorber
can be ascribed to peculiar velocities.) In fact, the DLA  absorbing cloud can
be regarded as an occulting disc, effective over a narrow  but significant
range of wavelengths around Lyman~$\alpha$, and enabling us to see any faint
object in front of it or beside it without interference from the QSO's light.

Given the large velocity offset between the absorber and emitter, it may be
that they form part of the same group, which could also include the QSO.
Further observations to search for other group members would be worthwhile.

It is possible that the QSO's radiation field may excite the emission
feature directly or stimulate it indirectly (e.g. by star formation),
rather than the feature being an independent self-luminous region. The
broad quasar Lyman~$\alpha$ emission line includes the wavelength of the
DLA. This implies a substantial flux from the QSO, both line and continuum,
at the DLA Lyman~$\alpha$ wavelength.  Although only a small number of DLAs
have been detected in emission, it is interesting to note the probable
over-representation of systems with DLA redshift close to the QSO's
emission redshift (PKS 0528$-$250; Q2059$-$360; Q0151+048A; see also M\o
ller et al. 1998). If confirmed by further cases, this would suggest the
nearby quasar does sometimes have a role in producing the emission from the
DLA region.

The low rate of detections of emission from DLAs has been remarked on many
times. We point out here that this result is closely related to the minimal
success of field searches for high redshift Lyman~$\alpha$ emitting
galaxies/protogalaxies (e.g. Thompson \& Djorgovski 1995).  Accepting that
the clouds which cause damped absorptions in QSO spectra are normally not
physically associated with the QSO, and based on the number statistics of
DLAs (Wolfe et al. 1995), it can be shown that a substantial fraction of
the sky has at least one potential high column density cloud at a redshift
which would place its Lyman~$\alpha$ absorption/emission in the visible.
Thus, if damped absorbers were frequently found in emission, the
corresponding clouds which do not happen to lie in front of a QSO would be
sufficiently common to have appeared in substantial numbers in the
blank-sky searches.  Even in our long-slit observations targeted at
Q2059$-$360, we could expect 7 to 14 such objects along the 3.1 arcminute
slit with Lyman~$\alpha$ in the observable redshift range of 2.457 to 2.832
(assuming each DLA system to have an angular size of 1 - 2$^{\prime\prime}$).
No other objects were found. We note, however, that establishing the
reality of an isolated emission feature at a position and wavelength not
indicated by any other data, would require a signal greater than we
observed towards Q2059$-$360. The above argument applies at the sensitivity
levels achieved by 4-m class telescopes; using the Keck II telescope Hu et
al. (1998) have shown that a substantial population of Lyman $\alpha$
emitting objects can indeed be found in the field.

\subsection{Relation of the Emission Feature to the QSO}
We consider three possibilities for the situation of the DLA emission
region with respect to the quasar, as shown in Figure \ref{fig.velpos2}b.

(i) Both the DLA absorption and the Lyman $\alpha$ emission could occur
within an object similar to the high redshift Lyman-break galaxies.  High
relative velocities between emission and (interstellar) absorption have
been demonstrated by Pettini et al. (1998). However, the existence and
survival of clouds with a DLA column density in such galaxies has yet to be
established.

(ii) It is possible that the absorber is well separated from the QSO. Since
the emission region cannot be directly behind the opaque DLA absorber, the
emission feature regions would not be physically associated with the
QSO. At most their emission could be enhanced by the influence of the
quasar's radiation field.  The emitter and absorber could be part of a
group or cluster. Clustering in redshift space has been recently reported
at these redshifts (Steidel et al. 1997b).

(iii) The third model in Figure \ref{fig.velpos2}b accounts for the near
equality of the emission feature and QSO redshifts (especially when the
latter is taken from the C~{\sc iv} line).  It interprets the emission
features as arising near the QSO, either from filaments of the quasar's
Narrow Line Region (NLR), or close companion(s) of the QSO.  The features
observed as spatially coincident with the QSO could be close beside it, at
a separation not resolved by the seeing disc of our observations. This
requires that the DLA absorbing cloud have a small physical size (or more
precisely, that the absorption have a sharp edge). The DLA could then
occult the QSO but leave visible the extended emission close beside it.
Their spatial extent is consistent with the emission features being due to
NLR filaments (Bremer et al. 1992, Heckmann et al. 1991). In this picture
it is possible for the DLA absorber to lie within the QSO host galaxy.

\section{Conclusion}

Our observations confirm that there is emission within the damped
Lyman~$\alpha$ absorption trough of Q2059$-$360, and we show that the
emission is extended both spatially and spectrally. The interpretation of
this emission is complicated by two factors: {\it (i)} the large velocity
offset between absorption and emission; {\it (ii)} the closeness of the QSO
emission redshift to the DLA emission feature.  The emission could be from
a high redshift galaxy which also contains the DLA absorbing cloud,
alternatively the emission feature could be due to object(s) in the
vicinity of the DLA absorber but distinct from it, or filaments or
companions of the QSO itself.

The nature of the system could be clarified by further observations, in
particular: {\it (i)} Detection of other emission lines (especially C~{\sc
iv}) and/or the continuum from the emission features; {\it (ii)}
Determination of a more accurate systemic velocity for the QSO, most likely
via detection of additional low-ionization emission lines; {\it (iii)} A
more comprehensive study of the metal lines in the DLA absorption system at
higher resolution. We will address these points in future work.

\subsection*{ACKNOWLEDGMENTS}

We thank the ESO Observing Programs Committee for allocation of telescope
time, and the ESO site staff for assistance. The Science Foundation for
Physics within the University of Sydney supported travel for JGR. We
also thank Dr. R. Hunstead for useful comments. We thank the referee for
comments which have improved the paper.


\begin{thebibliography}{99}
\bibitem{b1} Bremer M.N., Fabian A.C., Sargent W.L.W., Steidel C.C.,
 Boskenberg A., Johnstone R. M., 1992, MNRAS, 258 23P
\bibitem{b1} Charlot S., Fall S.M., 1991, ApJ, 378, 471
\bibitem{b1} Djorgovski S.G., Pahre M.A., Bechtold J., Elston R., 1996,
Nature, 382, 234
\bibitem{b1} Franceschini A., Gratton R., 1997, MNRAS, 286, 235
\bibitem{b1} Franx M., Illingworth G. D., Kelson D. D., van Dokkum P.
G., Tran K.-V., 1997, ApJ, 486, L75
\bibitem{b1} Fynbo J., M\o ller P. and Warren S.J., 1997,
Proc. conf. `Structure and Evolution of the Intergalactic Medium from
QSO Absorption Line Systems', ed. P. Petitjean, Paris, in press
\bibitem{b1} Gaskell C. M., 1982, ApJ, 263, 79
\bibitem{b1} Ge J., Bechtold J., Walker C., \& Black J. H., 1997, ApJ, 486, 727
\bibitem{b1} Giavalisco M., Steidel C.C., Macchetto F.D., 1996, ApJ, 470, 189
\bibitem{b1} Haehnelt M.G., Steinmetz M., Rauch M., 1998, ApJ, 495, 647
\bibitem{b1} Hamuy M., Suntzeff B.N., Heathcote S.R., Walker A.R.,
Gigoux P., Phillips M.M., 1994, PASP, 106, 566
\bibitem{b1} Hamann F., 1997, ApJS, 109, 279
\bibitem{b1} Hamann F., Barlow T.A., Junkkarinen V., Burbidge E.M.,
1997, ApJ, 478, 80
\bibitem{b1} Heckmann T.M., Lehnert M.D., Miley G.K., van Breugel W.,
1991, ApJ, 381, 373
\bibitem{b1} Hunstead R.W., Pettini M., Fletcher A.B., 1990, ApJ, 356, 23
\bibitem{b1} Hu E.M., Cowie L.L., Mcmahon R.G., 1998, ApJ, 502, L99
\bibitem{b1} Le Brun V., Bergeron J., Boiss\'{e} P., Deharveng J.M.,
1997, A\&A, 321, 733
\bibitem{b1} Leibundgut B., Robertson J.G., 1998, The Young Universe,
ed. S. D'Odorico, E. Giallongo, ASP Conf. Ser. 146, 186
\bibitem{b1} Lequeux J., Kunth D., Mas-Hesse J.M., Sargent W.L.W., 1995,
A\&A, 301, 18
\bibitem{b1} Lowenthal J.D., Koo D.C., Guzman R., Gallego J., Phillips
A.C., Faber S.M., Vogt N.P., Illingworth G.D., Gronwall C., 1997,
ApJ, 481, 673
\bibitem{b1} Lowenthal J.D., Hogan C.J., Green R.F., Woodgate B., Caulet
A., Brown L., Bechtold J., 1995, ApJ, 451, 484
\bibitem{b1} Lu L., Sargent W.L.W., Barlow T.A., Churchill C.W.,
Vogt S.S., 1996, ApJS, 107, 475
\bibitem{b1} Lu L., Sargent W.L.W., \& Barlow T.A., 1997, ApJ, 484, L131
\bibitem{b1} Mar D.P., Bailey, G., 1995, PASA, 12, 239
\bibitem{b1} M\o ller P., Warren S.J., 1993, A\&A, 270, 43
\bibitem{b1} M\o ller P., Warren S.J., \& Fynbo J. U., 1998, A\&A, 330, 19
\bibitem{b1} Osterbrock D.E., 1989, Astrophysics of Gaseous Nebulae \&
Active Galactic Nuclei, (Mill Valley: University Science Books)
\bibitem{b1} Pascarelle S.M., Windhorst R.A., Keel W.C., Odewahn S.C.,
1996, Nature, 383, 45
\bibitem{b1} Petitjean P., Bergeron J., 1994, A\&A, 283, 759
\bibitem{b1} Pettini M., Hunstead R.W., King D.L., Smith L.J., 1995, QSO
Absorption Lines, ed. G. Meylan, (Berlin: Springer), 55
\bibitem{b1} Pettini M., Smith L.J., King D.L., Hunstead R.W., 1997,
ApJ, 486, 665
\bibitem{b1} Pettini M., Kellogg M., Steidel C.C., Dickinson M., Adelberger
K.L., Giavalisco M., 1998 preprint (astro-ph 9806219)
\bibitem{b1} Prochaska J.X., Wolfe A.M., 1996, ApJ, 470, 403
\bibitem{b1} Prochaska J.X., Wolfe A.M., 1997, ApJ, 487, 73
\bibitem{b1} Steidel C.C., Giavalisco M., Pettini M., Dickinson M.,
Adelberger K.L., 1996, ApJ, 462, L17
\bibitem{b1} Steidel C.C., Dickinson M., Meyer D.M., Adelberger K.L.,
Sembach K.R., 1997a, ApJ, 480, 568
\bibitem{b1} Steidel C.C., Adelberger K.L, Dickinson M., Giavalisco M.,
Pettini M., Kellogg M., 1997b, ApJ, 492, 428
\bibitem{b1} Storrie-Lombardi L.J., McMahon R.G., Irwin M.J., 1996,
MNRAS, 283, L79
\bibitem{b1} Thompson D., Djorgovski S.G., 1995, AJ, 110, 982
\bibitem{b1} Tytler D., Fan X.-M., 1992, ApJS, 79, 1
\bibitem{b1} Vogel S., Reimers D., 1995, A\&A, 294, 377
\bibitem{b1} Warren S.J., Hewett P.C., Osmer P.S., 1991, ApJS, 76, 23
\bibitem{b1} Warren S.J., M\o ller P., 1996, A\&A, 311, 25
\bibitem{b1} Wolfe A.M., Turnshek D.A., Lanzetta K.M., Oke
J.B., 1992, ApJ, 385, 151
\bibitem{b1} Wolfe A.M., Lanzetta K.M., Foltz C.B., Chaffee
F.H., 1995, ApJ, 454, 698
\bibitem{b1} Womble D.S., Sargent W.L.W., Lyons R.S., 1996 `Cold Gas at
High Redshift', ed. M.N. Bremer et al., Dordrecht: Kluwer, 249
\bibitem{b1} Zheng W., Kriss G.A., Telfer R.C., Grimes J.P., Davidsen
A.F., 1997, ApJ, 475, 469
\end{thebibliography}
\end{document}